\documentclass[9pt,twocolumn,twoside]{opticajnl}
\journal{opticajournal} % use for journal or Optica Open submissions

\setboolean{shortarticle}{true}
\usepackage{lineno}
\usepackage{physics}
\title{Errors in quantum state identification with ultrashort pulses}

\author[1,*]{Joscelyn van der Veen}
\author[1]{Daniel F.V. James}

\affil[1]{
  Department of Physics, University of Toronto, 60 St. George St., Toronto, Canada}

\affil[*]{joscelyn.vanderveen@utoronto.ca}

\begin{abstract}
From the nonclassicality of photon superbunching to the basic property of intensity, we characterize light with correlation and coherence functions. The correlation functions for nonstationary sources, such as short pulses, depend deterministically on the time dependent shape of the field and change the probability of the ensemble in time. We start from the fundamental principles of detection and show how nonstationary fields result in a coherence function that depends not only on the statistics of the field but also on the field shape. This means that observations of bunching and superbunching with ultrafast sources are not indicative of the quantum state of the field.
\end{abstract}

\setboolean{displaycopyright}{false} % Do not include copyright or licensing information in submission.

\begin{document}

\maketitle

The probabilistic nature of quantum mechanics means we cannot measure a state but we can deduce a likely state from many measurements \cite{Smithey1993,James2001}. With light, we deduce the state from its interference which allows us to measure the varying orders of coherence functions. One of the most useful coherence functions for identifying common states of light is the second order coherence function, the $g^{(2)}$ \cite{Glauber1963}. A single photon state is often defined by a $g^{(2)}$ of zero \cite{Gerry2004}, $g^{(2)}\approx2$ is used to characterize thermal (chaotic) sources and determine their coherence length \cite{Loudon2000}, and superbunching $g^{(2)}>2$ is supposed to indicate nonclassicality \cite{Theidel2024}. The second order coherence function is currently being used to characterize not only the stationary light sources for which it was originally derived, but also nonstationary light sources such as short pulses. A $g^{(2)}$ close to two is used as evidence of the thermal distribution of spontaneous parametric downconversion \cite{Tapster1998,Blauensteiner2009} and very recently superbunching in high harmonic generation was used as evidence of its nonclassicality. However, we show that the deterministic temporal shape of nonstationary light changes the averaging in coherence measurements and leads to larger expected $g^{(2)}$ values regardless of the statistics of the light source.

% The areas of ultrafast and quantum optics are generally described with different mathematical languages: ultrafast optics dealing with the deterministic time dependence of the field and quantum optics dealing with single frequency states. Recently, there has been much interest in the quantum state of ultrashort pulses with the improvements predicted for high harmonic generation (HHG) with squeezed light \cite{Gorlach2023}. The evidence for the quantum nature of both the bright squeezed vacuum used to drive HHG in \cite{Rasputnyi2024} and for the harmonics in HHG themselves \cite{Theidel2024} is related to superbunching, which is a light with a larger second order coherence ($g^{(2)}>2$) than chaotic light ($g^{(2)}=2$). We show that the apparent superbunching is in fact due to the deterministic temporal shape of short pulses in typical $g^{(2)}$ measurement, which disguises the effect of quantum correlations.

Any photon detector can be broken down into single atom-like photon detectors that are exposed to the electromagnetic field from some time $t_0$ to some later time $t$ \cite{Glauber2006}. The transition probability of a single atom-like detector being excited from a ground to excited state in an exposure time $t_0$ to $t$ for a broadband detector is,

\begin{equation}
    p(t)=s\int_{t_0}^t G^{(1)}(\boldsymbol{r},t') \dd{t'}
\end{equation}

\noindent where $s$ is the sensitivity factor dependent only on the properties of the detector and

\begin{equation}
    G^{(1)}(\boldsymbol{r},t')=\Tr{\hat{\rho} \hat{\boldsymbol{E}}^{(-)}(\boldsymbol{r},t)\hat{\boldsymbol{E}}^{(+}(\boldsymbol{r},t))}
    \label{eq:G1}
\end{equation}

\noindent is the first order correlation function of the field.

If we know the density matrix $\hat{\rho}$ of the electric field, then we can determine the probability of excitation of our detection with a given exposure time. However, this requires us to know the correlation function Eq. \ref{eq:G1}. The correlation function is an ensemble average so it must be determined over many measurements of the field. If we consider a detector being illuminated with a field whose density matrix remains the same, also known as a stationary field \cite{Mandel_Wolf_1995}, then the detector can be excited repeatedly taking some amount of time as given by the probability distribution. The rate at which this excitation occurs is the derivative of the probability distribution and we call it a countrate or an intensity,

\begin{equation}
    I(\boldsymbol{r},t)\propto G^{(1)}(\boldsymbol{r},t)=\Tr{\hat{\rho} \hat{\boldsymbol{E}}^{(-)}(\boldsymbol{r},t)\hat{\boldsymbol{E}}^{(+)}(\boldsymbol{r},t))}
\end{equation}

A nonstationary field, such as a pulse train, essentially acts as a shutter. This means we can no longer talk about a countrate in the same way because we have definite time periods during which the detector is exposed to more than vacuum fluctuations of the field. Let us therefore consider intensity from a statistical perspective.

To detect a field, a single atom-like detector must be excited and trigger some process before resetting to be able to detect again. An individual photodetection thus only has two results at any given spacial point and detector reset time window: detection (one) or no detection (zero).

Detections and no detections occur probabilistically so to find the average number of detections we must take an ensemble of measurements. The elements of the ensemble could be spatially or temporally separated but they still must occur during some non-infinitesimal time window. For example, if we had a detector composed of a grid of $N$ atoms then we could define a field intensity for a given time window as the proportion of atoms that were excited during that time window. We could similarly define a field intensity temporally by allowing the detector to reset after an excitation and measuring whether there is a detection during a series of time windows. To measure the field rather than the detector limitations, we require a low enough probability of excitation that the detector will not always click during some time window after resetting. With either spatial or temporal averaging, we are determining the intensity based on a number of detections that occur during some imposed time window so the intensity is a photon rate.

Even if we define intensity based on an average of detection and no detection, we can partition these detections arbitrarily. Consider a set of $N$ detector measurements $\{x_i\}$ taken either spatially or temporally. Each measurement $x_i$ can take the value one or zero, corresponding to detection or no detection respectively, and the intensity can be given by the average $\frac{1}{N}\sum_{i=1}^Nx_i$. However, it could equivalently be given by an average of averages 

\begin{equation}
    I=\frac{1}{a}\sum_{j=1}^a\frac{1}{N/a}\sum_{i=1+(j-1)N/a}^{jN/a}x_i
    \label{eq:G1-average}
\end{equation} 

For a stationary field, this means we retrieve the intensity as a photon countrate that we can express without the actual detection or no detection time window. For example, if we average a stationary field temporally with an arbitrary window, we could take $a$ in Eq. \ref{eq:G1-average} to be the number of seconds we measured and the intensity would then be an average of the average number of photons each second, regardless of the time window for detection or no detection.

To understand the intensity for nonstationary light, let us begin by naively assuming the intensity is still proportional to the first order correlation function. We can expand the positive frequency electric field operator in terms of temporal modes \cite{Titulaer1966},

\begin{equation}
    \hat{E}^{(+)}(\boldsymbol{r},t)=\sum_l\boldsymbol{v}_l(\boldsymbol{r},t)\hat{b}_l
\end{equation}

\noindent where 

\begin{equation}
    \boldsymbol{v}_l(\boldsymbol{r},t)\equiv i\sum_k\gamma_{lk}\sqrt{\frac{\hbar\omega_k}{2\epsilon_0V}}\boldsymbol{u}_k(\boldsymbol{r})e^{-i\omega_kt} 
    \label{eq:vl}
\end{equation}

\noindent if $\gamma_{lk}$ are the elements of a unitary matrix.

The first order correlation function for a field in a single temporal mode, say the $0$th mode, is then,

\begin{equation}
    G^{(1)}(\boldsymbol{r},t)=\left|\boldsymbol{v}_0(\boldsymbol{r},t)\right|^2\Tr{\hat{\rho_0}\hat{b}_0^\dagger\hat{b}_0}
\end{equation}

The time dependence of the field is entirely deterministic and separate from the state of the field. This means the temporal mode shape Eq. \ref{eq:vl} acts exactly as we would expect for a shutter: it only allows detections during a certain time period and does not affect the statistics of measurements during that time period. So long as we include the vacuum states of the remaining temporal modes when considering the uncertainty in the intensity, it is reasonable to still consider the intensity of a nonstationary field to be proportional to the first order correlation function.

In our detection/no detection definition of intensity, the ensemble of $N$ measurements $\{x_i\}$ now do not have consistent probabilities of detection when we average temporally. For $N$ measurements taken one after another, only $M$ of those measurements will occur while there is some amplitude of the field. The other $N-M$ measurements will be zero aside from counts arising from the uncertainty of the vacuum state. If we order our data with the $M$ measurements first, the intensity is then given by $\frac{1}{N}\sum_1^Mx_i=R_I\frac{1}{M}\sum_1^Mx_i$ where $\frac{1}{M}\sum_1^Mx_i$ is the intensity of the $M$ measurements and we define the quantity $R_I\equiv M/N$ as the intensity ratio. The intensity ratio is clearly proportional to the pulse width for a pulse train but also depends on factors such as the repetition rate and the detector reset time. 

For the first order correlation function, we only expect the intensity to be proportional to the first order correlation function so the $R_I$ factor doesn't matter. However, let us now consider higher order correlation functions,

\begin{multline}
    G^{(n)}(\boldsymbol{r}_1,\ldots\boldsymbol{r}_n;t_1,\ldots,t_n) \\
    \\
    =\Tr\left\{\hat{\rho}\hat{\boldsymbol{E}}^{(-)}(\boldsymbol{r}_1,t_1)\ldots\hat{\boldsymbol{E}}^{(-)}(\boldsymbol{r}_n,t_n)\right. \\
    \left.\hat{\boldsymbol{E}}^{(+)}(\boldsymbol{r}_n,t_n)\ldots\hat{\boldsymbol{E}}^{(+)}(\boldsymbol{r}_1,t_1)\right\}
\end{multline}

\noindent and coherence functions,

\begin{equation}
    g^{(n)}(\boldsymbol{r}_1,\ldots\boldsymbol{r}_n;t_1,\ldots,t_n)=\frac{G^{(n)}(\boldsymbol{r}_1,\ldots\boldsymbol{r}_n;t_1,\ldots,t_n)}{G^{(1)}(\boldsymbol{r}_1,t_1)\ldots G^{(1)}(\boldsymbol{r}_n,t_n)}
    \label{eq:gn}
\end{equation}

In a single temporal mode, the coherence function is not explicitly dependent on time,

\begin{equation}
    g^{(n)}\equiv \frac{\expval{\left(\hat{b}_0^\dagger\right)^n\left( \hat{b}_0\right)^n}}{\expval{\hat{b}_0^\dagger \hat{b}_0}^n}
\end{equation}

However, correlation functions remain ensemble averages and each measurement of the ensemble takes a non-infinitessimal time. 

From a statistical perspective, a single measurement of an $n$th order correlation can only be a one if there is a detection on $n$ detectors during a time window or zero if there is no detection on any of the $n$ detectors during that time window. For the second order correlation function, this would be two sets of data $\{x_i\}$ and $\{y_i\}$ that we multiply elementwise $\{x_iy_i\}$. The elements $x_i$ and $y_i$ are either one or zero and must be measured during the same time window. When considering higher order correlations, we can similarly partition the average into correlations with an average intensity greater than one that we then average since,

\begin{multline}
    \frac{1}{N}\sum_{i=1}^N\left(x^{(1)}_ix^{(2)}_i\ldots x^{(n)}_i\right) \\
    =\frac{1}{a}\sum_{j=1}^a\frac{1}{N/a}\sum_{i=1+(j-1)N/a}^{jN/a}\left(x^{(1)}_ix^{(2)}_i\ldots x^{(n)}_i\right)
\end{multline}

Importantly, the partitioning of the average still requires the elements $x^{(j)}_i$ for the $j=[1,n]$ measurements from the same time window to be multiplied together. Therefore, regardless of how we partition, the $n$th order correlation function for a nonstationary field averaged temporally will only consider the $M$ detections when the field has amplitude.

For example, consider a typical measurement of second order coherence, such as a Hanbury-Brown-Twiss (HBT) interferometer, where we measure the correlations at two single click detectors. If we consider detections at both detectors during $N$ time windows for an ensemble of measurements $x_i$ and $y_i$, 

\begin{equation}
    g^{(2)}=\frac{\frac{1}{N}\sum_{i=1}^Mx_iy_i}{\frac{1}{N}\sum_{i=1}^Mx_i\frac{1}{N}\sum_{i=1}^My_i}=\frac{1}{R_I}\frac{\frac{1}{M}\sum_{i=1}^Mx_iy_i}{\frac{1}{M}\sum_{i=1}^Mx_i\frac{1}{M}\sum_{i=1}^My_i}
\end{equation}

Thus, the $g^{(2)}$ is inversely proportional to the intensity ratio $R_I$. This equation is the main result of the paper since $R_I<1$ and the intensity ratio will thus lead to large $g^{(2)}$ values even for statistically independent (coherent state) sources.

This has already been shown both experimentally \cite{Riedmatten2004} and theoretically \cite{Ou1999} for $g^{(2)}$ measurements of one arm of a spontaneous parametric downconversion (SPDC) source. While these works did not explicitly acknowledge the result being due to the determinism of a time dependent field shape, they both use that assumption to show that the $g^{(2)}$ is inversely proportional to the pulse duration. We also see an increase in the $g^{(2)}$ for decreasing photon number in the measurement of harmonics from high harmonic generation with an HBT interferometer \cite{Theidel2024}.

We can verify the effect of $R_I$ on the $g^{(2)}$ by measuring a $g^{(2)}$ with an HBT interferometer for a wider variety of short pulse sources. As we decrease the pulse width but maintain the same average photon number, we expect to see the $g^{(2)}$ to increase.

Further, to measure a $g^{(2)}$ for a short pulse source that describes the statistics analogously to a stationary $g^{(2)}$, we can see that we should instead measure the statistical correlations over an ensemble of pairs of detectors. We can also see the difficulty in this type of measurement because it would require the time dependent pulse shape to be split with equal amplitude going to each pair of detectors.

We may also consider measuring detections and no detections only during the nonzero amplitude of a pulse. However, this reduces us to an ensemble of $M$ measurements. For any pulse whose duration is short enough for only one possible detection per detector reset time, we are reduced to $M=1$, or an ensemble of one measurement. When we are reduced to one measurement per pulse, we can only get an ensemble by considering multiple pulses. Pulse to pulse correlations are an interesting quantity to measure, but they are not analogous to stationary correlation functions. The pulse to pulse coherence would therefore not identify the quantum state of the field as we define states under canonical quantization.

We have shown that from fundamental statistics, the second order coherence function measured with temporal averaging changes due to the deterministic pulse shape regardless of the state of the field. Further, we have proposed experiments to confirm this behaviour of the $g^{(2)}$ and how to measure a $g^{(2)}$ for a nonstationary source that gives analogous statistical information to a measurement with a stationary source. This not only improves our understanding of correlations in short pulses but also shows there may be errors in the quantum states assigned to short pulse fields. A satisfactory means to characterize the quantum nature of light for nonstationary fields requires simultaneous ensemble averaging and measurements of such are currently lacking. 

\begin{backmatter}
\bmsection{Funding} National Science and Engineering Research Council of Canada (579228 - 2023)

\bmsection{Acknowledgment} We wish to acknowledge Donna Strickland for useful conversation that inspired this paper.

\bmsection{Disclosures} The authors declare no conflicts of interest.

\bmsection{Data availability} No data were generated or analyzed in the presented research.

\end{backmatter}

% Bibliography
\bibliography{sample}

% Full bibliography added automatically for Optics Letters submissions; the following line will simply be ignored if submitting to other journals.
% Note that this extra page will not count against page length
\bibliographyfullrefs{sample}

\end{document}